\documentclass[12pt]{article}
\begin{document}
\begin{center}
\Large{{\bf Chiral Multiplets in Hadronic Physics}}
\end{center}
\vspace*{0.5cm}
\begin{center}
\textbf {A. E. Inopin}
\end{center}
\vspace*{0.5cm}
\begin{center}
\large{Kharkov National University, Department of Experimental
Nuclear Physics, Svobody Sq. 4, 61077, Kharkov, Ukraine}
\end{center}
\vspace*{0.5cm}

We consider the multiplet structures, appearing in hadronic
physics. The basic types of the multiplets revealed are parity
doublets, chiral multiplets, and multispin-parity clusters. We
elucidate a new type of isospin supermultiplets in the baryon
sector. The creation of parity doublets, chiral multiplets and
their interrelation with multispin-parity clusters has been
uncovered. The role of chiral symmetry breaking and restoration
in the formation of the hadronic spectra has been scrutinized.

\vskip 0.8 cm
{\bf \ \ \ \ \ \ \ \
\ \ \ \ \ \ \ \ \ \ \ \ \ \ \ \ \ \ \ \ \ \ \ \ \ \ I.
INTRODUCTION} \vskip 0.7cm

Chiral symmetry plays a vital role in strong interaction
physics. Since its genesis, the quark model has had to deal with
the different facets of the quarks. From the field-theoretical
point of view, bare quarks are massless or have very small mass
of a few MeV (u,d-quarks). These are also called the "current
quarks". The current quark masses are determined by the
QCD-sumrules. On the other hand there exists a big class of quark
models - potential, etc., which deals with so-called "constituent
quarks" with masses of about 1/3 of the nucleon's mass.

The idea behind chiral models is this lightness of the light
quarks compared to typical hadronic scales (take roughly 1GeV or
the nucleon mass). One starts with massless quarks and realized
that in this so-called chiral limit, the QCD Lagrangian shows an
additional symmetry: $SU(2)_L\times SU(2)_R$. This symmetry is
broken only by the very small u,d quark masses. However this
symmetry is not directly observed in the world - it is {\it
hidden}, i.e. {\it spontaneously broken symmetry} (SBS)
\cite{1}.

Creation of the universe in cosmology, generation of gauge bosons,
Higgs particles, t'Hooft monopoles, instantons and solitons in
high energy elementary particle physics, and phenomena of laser,
superradiance, superconductivity, superfluidity, and phase
transition, are all well understood in terms of SBS. Those
phenomena are all known as {\it order-creating} phenomena in
nature.

Symmetry and order are two mutually complementary concepts; when
we have rotational symmetry, there is no particular direction
singled out as being different. When we notify a specific
direction, there is directional order, and the rotational
symmetry should be lost to create such an order. The mechanism of
SBS was first demonstrated theoretically in quantum field theory.
In the system of {\it infinitely many} degrees of freedom
described by the Lagrangian manifesting the rotational symmetry,
only one state is chosen spontaneously among the infinitely
degenerate ground state of the system and the rotational symmetry
of the system is broken without recourse to any external
environment. Although SBS is a mechanism characteristic to the
system of infinitely many degrees of freedom described by quantum
field theory, it can equally be applied to the system of finite
but many degrees of freedom. "Spontaneous" means the following
fact. The rotational symmetry is broken not by imposing certain
external force, but by the fact that the system itself chooses
one and only one ground state among the infinitely many possible
ground states and the transition of the chosen ground state into
other ones cannot be realized. Due to a simple calculation in
quantum field theory, the more degrees of freedom of the system
becomes, the more stable the spontaneously chosen ground state is.
Thus in the case of infinitely many degrees of freedom, the ground
state with SBS becomes highly stable but manifests the broken
rotational symmetry.

The idea that the fundamental strong interaction theory should
possess an approximate $SU(2)_L\times SU(2)_R$ or ($SU(3)_L\times
SU(3)_R$) chiral symmetry dates back to the 1960s. One of the
most important insights from this was that this symmetry must be
spontaneously broken in the vacuum (i.e. realized in the
Nambu-Goldstone mode). The most important early arguments were :
i) the absence of parity doublets (PD) in the hadron spectrum (if
the chiral symmetry were realized in the Wigner-Weyl mode - i.e.
if the vacuum were trivial - then the hadron spectrum would have
to reveal the multiplets of the chiral group which are manifested
as parity doublets); (ii) the exceptionally low mass of pions,
which are taken to be pseudo-Goldstone bosons associated with the
{\it spontaneously broken axial symmetry}.

During the last few years it has been realized that the spectra
of baryon and mesonic resonances exhibit few startling features.
First of all it is a clustering, or multiplet structure of the
experimental spectra. More precisely, the majority of the hadronic
resonances are grouped into some kind of clusters, and we term
them as "chiral multiplets" (the meaning of this term will become
clear later). Prof. H\"{o}ehler was probably the first who
realized a cluster structure of the baryonic spectra, i.e. baryon
resonance should {\it not} be treated as an individual states,
but rather as a set of clusters, "H\"{o}ehler clusters" (or a
H\"{o}ehler poles) \cite{2}. Another startling feature is the
occurrence of the parity doublets in the hadronic spectra. By this
we mean the mass degeneracy among physical resonances with the
same total angular momentum {\it J}, but the opposite {\it P}-
parities \cite{3}.

In what follows we will try to uncover the origins of these truly
remarkable phenomena. We will also establish their similarities
and differences, and what are the real dynamics hidden behind all
of this. We will scrutinize what is the role of the chiral
symmetry breaking and restoration in the formation of parity
doublets and chiral multiplets. These mechanisms and
manifestation of chiral symmetry are different for baryons and
mesons.

We will analyze the whole baryon spectra and will construct
parity doublets and chiral multiplets. In so doing, we analyze a
few different models of this phenomena: Klempt, Inopin,
Cohen-Glozman, Hatsuda, Kirchbach. We discover {\it duality}
between the different models and how it sheds light on this
problem. New types of symmetries describing this multifaceted
phenomena will be uncovered.
\vskip 0.7cm

{\bf \ \ \ \ \ \ \ \ \ \ \ \ \ \ \ \ \ \ \ \ \ \ \ \
\ \ \ \ \ \ \ \ \ \ \ II. Chiral Multiplets}
\vskip 0.7cm

We will attack this issue from different directions
simultaneously. On one hand, we will consider experimental
spectra and look for the emergence of clusters. On the other
hand, the block of modern quark models will be presented, where
authors show the multiplet structure from the dynamical and
symmetry arguments.
\vskip 1.0cm
\newpage
{\bf \ \ \ \ \ \ \ \ \ \ \ \ \ \ \ \ \ \ \ \ \ \ \ \ \ \ \ \ \ \ \
\ \ \ \ a) \underline {Cohen-Glozman model}}

\vskip 0.3cm

The authors consider only $N$, $\Delta$ spectra \cite{4}.
They suggested that the parity doublet structure seen in the
spectrum of highly excited baryons may be due to effective chiral
restoration for these states. The authors have argued how the
idea of chiral symmetry restoration high in the spectrum is
consistent with the concept of quark-hadron duality (QHD). If
chiral symmetry is effectively restored for high-lying states,
then the baryons should fall into representation of
$SU(2)_L\times SU(2)_R$ that are compatible with the given
parity of the states-the parity-chiral multiplets. The authors
classify all possible parity-chiral multiplets: (i) (1/2,
0)$\oplus$(0,1/2) that contain parity doublet for the nucleon
spectrum, (ii) (3/2, 0)$\oplus$(0,3/2) consists of the parity
doublet for delta spectrum, (iii) (1/2, 1)$\oplus$(1,1/2) contains
one parity doublet in the nucleon spectrum and one parity
doublet in the delta spectrum of the same spins that are
degenerate in mass. They show that the available spectroscopic
data for non-strange baryons in the $\sim 2GeV$ range are
consistent with all possibilities, but the approximate degeneracy
of parity doublets in nucleon and delta spectra support the
latter possibility with excited baryons approximately falling
into (1/2, 1)$\oplus$(1,1/2) representation of $SU(2)_L\times
SU(2)_R$ with approximate degeneracy between positive and negative
parity $N$ and $\Delta$ resonances of the same spin \cite{4}.

The main conjecture here is that high enough in the spectrum,
chiral symmetry is restored and this is {\it precisely the reason}
of the parity doublet occurrence and chiral multiplets creation.
This actually has not been proven. Another moment is the choice
of 2 GeV for the role of the high mass region's threshold. Beside
the fact that it is hardly possible to draw such a boundary line
between the low- and high-mass region, there are simply {\it too
many baryon resonances} (BR) above 2 GeV and they stretch up to
4.1 GeV \cite{5}. So, 2 GeV happens to be just in the middle of
$N$ $\Delta$ spectrum. Let's present the major result of
Cohen-Glozman model in a Table 1.

As we can see, some chiral multiplets have a big mass spread and
hardly could be considered as a degenerate set of physical
states: {\it \textbf{J}}$=1/2$, $\delta (spread)$ = 200 MeV; {\it
\textbf{J}}$=3/2$, $\delta$ = 180 MeV; {\it \textbf{J}}$=5/2$,
$\delta$ = 295 MeV;

\begin{table}[ht]
\caption{\textbf{Chiral Multiplets in Cohen-Glozman Model}}
\begin{center}
\begin{tabular}{| c | c | c | c | c |}
\hline
{\it \textbf{J}}$=1/2$ & $N^+(2100)$ & $N^-(2090)$ & $\Delta^+(1910)$ & $\Delta^-(1900)$ \\
\hline
{\it \textbf{J}}$=3/2$& $N^+(1900)$ & $N^-(2080)$ & $\Delta^+(1920)$ & $\Delta^-(1940)$ \\
\hline
{\it \textbf{J}}$=5/2$ & $N^+(2000)$ & $N^-(2200)$ & $\Delta^+(1905)$ & $\Delta^-(1930)$ \\
\hline
{\it \textbf{J}}$=7/2$& $N^+(1990)$ & $N^-(2190)$ & $\Delta^+(1950)$ & $\Delta^-(2200)$ \\
\hline
{\it \textbf{J}}$=9/2$ & $N^+(2220)$ & $N^-(2250)$ & $\Delta^+(2300)$ & $\Delta^-(2400)$ \\
\hline
{\it \textbf{J}}$=11/2$ & ? & $N^-(2600)$ & $\Delta^+(2420)$ & ? \\
\hline
{\it \textbf{J}}$=13/2$ & $N^+(2700)$ & ? & ? & $\Delta^-(2750)$ \\
\hline
{\it \textbf{J}}$=15/2$ & ? & ? & $\Delta^+(2950)$ & ? \\
\hline
\end{tabular}
\end{center}
\end{table}

{\it \textbf{J}}$=7/2$, $\delta $ = 250 MeV; {\it
\textbf{J}}$=9/2$, $\delta$ = 180 MeV; {\it \textbf{J}}$=11/2$,
$\delta$ = 180 MeV (unfilled);{\it \textbf{J}}$=13/2$, $\delta $ =
50 MeV (unfilled); {\it \textbf{J}}$=15/2$, unfilled.

So for the filled quartets, the spread varies in a range 180 MeV
$<\delta<295$ MeV, with $<\delta>=221$ MeV, which is much greater
that the pion mass. This could mean possible transitions between
the quartet members with pion emission.

Recently, the SAPHIR Collaboration (ELSA) tentatively discovered
a few $N^*$, precisely: $N_{{1/2}^-}(1897)$, $N_{{1/2}^+}(1986)$,
and $N_{{3/2}^-}(1895)$ \cite{5}. We can try to insert this
triplet into Cohen-Glozman model and see if the chiral multiplets
get narrower. The authors \cite{4} also neglect $N$, $\Delta
(\sim3000$ region) data from the PDG2002 which has been well-known
for years. Namely, we are talking about the following class of
resonances, obtained by Koch and Hendry:

\begin{table}[ht]
\caption{\textbf{N,$\Delta \sim 3000$ Region, from PDG2002}}
\begin{center}
\begin{tabular}{| c | c | c | c | c |}
\hline
{\it \textbf{J}}$=11/2$ & & & & $\Delta^-(2850)$ \\
\hline
{\it \textbf{J}}$=13/2$ & & & $\Delta^+(3200)$ & \\
\hline
{\it \textbf{J}}$=15/2$ & $N^-(3100)$ & $N^-(3500)$ & & \\
\hline
{\it \textbf{J}}$=17/2$ & $N^+(3500)$ & $N^+(3800)$ & & $\Delta^-(3300)$ \\
\hline
{\it \textbf{J}}$=19/2$ & $N^-(3750)$ & $N^-(4100)$ & $\Delta^+(3500)$ & $\Delta^+(3700)$ \\
\hline
{\it \textbf{J}}$=21/2$ & & & &$\Delta^-(4100)$ \\
\hline
\end{tabular}
\end{center}
\end{table}

So, we have in total 3+12=15 extra BR's which we will include in
this analysis. But that's not all. Recently we have developed a
potential quark model, which could describe $N$, $\Delta$ and
strange resonances with both small $J$, $M$ and large $J$, $M$. In
particular this model predicts a whole series of high-lying $N$,
$\Delta$ resonances, represented in Table 3 \cite{6},\cite{7}.
After much experimenting with Tables 1,2 and 3 and the SAPHIR
states we can suggest a modified Cohen-Glozman chiral multiplets,
presented in Table 4.

One can see that {\it \textbf{J}}$=1/2$ quartet now has 89 MeV
mass spread (it was 180 MeV). This is significant step forward in
getting really narrow multiplets! We have marked all of the new
entries by an asterisk and in color in Table 4. Another
interesting feature of this modification is that some multiplets
have gotten filled in more, i.e.: {\it \textbf{J}}$=11/2$ has 4
states instead of 2; {\it \textbf{J}}$=13/2$ has 4 states instead
of 2; {\it \textbf{J}}$=15/2$ has 4 quite close states instead of
just one. The third startling feature is the emergence of a new
sector with {\it \textbf{J}}$=17/2$, $19/2$ and $21/2$.

\vskip 1 cm

\begin{table}[ht]
\caption{\textbf{High-Lying N,$\Delta$ states, predicted in our
potential model \cite{6},\cite{7}}}
\begin{center}
\begin{tabular}{| c | c | c |}
\hline
$N_{{11/2}^+}(2390)$ & $N_{{13/2}^-}(2900)$ & $N_{{15/2}^+}(3000)$ \\
\hline
$N_{{17/2}^-}(3450)$ & $N_{{19/2}^+}(3450)$ & $N_{{21/2}^-}(3950)$ \\
\hline
$\Delta_{{15/2}^-}(3480)$& $\Delta_{{17/2}^+}(3490)$ & $\Delta_{{19/2}^-}(4000)$ \\
\hline

\end{tabular}
\end{center}
\end{table}

Originally Cohen-Glozman has 25 states, compared to the modified
scheme with 42 states, which is about a 40 \%
difference . Table 4 is filled up to 95\%.

\begin{table}[ht]
\caption{\textbf{Modified Cohen-Glozman-Inopin Multiplets}}
\begin{center}
\begin{tabular}{| c | c | c | c | c |}
\hline
\large{ {\it \textbf{J}}$=1/2$} & \large{ {$N^{*+}(1986)$}} &
\large{ {$N^{*-}(1897)$}} & \large{$\Delta^+(1910)$} &
\large{$\Delta^-(1900)$} \\

\hline
\large{ {\it \textbf{J}}$=3/2$}& \large{$N^+(1900)$} &
\large{ {$N^{*-}(1895)$}} & \large{$\Delta^+(1920)$} &
\large{$\Delta^-(1940)$} \\
\hline
\large{ {\it \textbf{J}}$=5/2$}& \large{$N^+(2000)$} & \large{$N^-(2200)$} &
\large{$\Delta^+(1905)$} & \large{$\Delta^-(1930)$} \\
\hline
\large{ {\it \textbf{J}}$=7/2$} & \large{$N^+(1990)$} & \large{$N^-(2190)$} &
\large{$\Delta^+(1950)$} & \large{$\Delta^-(2200)$} \\
\hline
\large{ {\it \textbf{J}}$=9/2$} & \large{$N^+(2220)$} & \large{$N^-(2250)$} &
\large{$\Delta^+(2300)$} & \large{$\Delta^-(2400)$} \\
\hline
\large{ {\it \textbf{J}}$=11/2$} &\large{ {$N^+(2390)$}} &
\large{$N^-(2600)$} & \large{$\Delta^+(2420)$} &
\large{ {$\Delta^{*-}(2850)$}} \\
\hline
\large{ {\it \textbf{J}}$=13/2$} & \large{$N^+(2700)$} &
\large{ {$N^-(2900)$}}& \large{ {$\Delta^{*+}(3200)$}}
& \large{$\Delta^-(2750)$} \\
\hline
\large{ {\it \textbf{J}}$=15/2$}& \large{ {$N^+(3000)$}}&
\large{ {$N^{*-}(3100)$}} & \large{$\Delta^+(2950)$} &
\large{ {$\Delta^-(3480)$}} \\
\hline
\large{ {\it \textbf{J}}$=17/2$} & \large{ {$N^{*+}(3500)$}} &
\large{ {$N^-(3450)$}} & \large{ {$\Delta^+(3490)$}} &
\large{ {$\Delta^{*-}(3300)$}} \\
\hline
\large{ {\it \textbf{J}}$=19/2$} & \large{ {$N^{+}(3500)$}} &
\large{ {$N^{*-}(3750)$}} &
\large{ {$\Delta^{*+}(3500)$}} &
\large{ {$\Delta^-(4000)$}} \\
\hline
\large{ {\it \textbf{J}}$=21/2$} & \large{?} & \large{ {$N^-(3950)$}}
& \large{?} & \large{ {$\Delta^{*-}(4100)$}} \\
\hline
\end{tabular}
\end{center}
\end{table}

Let's return again to the original Cohen-Glozman scheme (Table 1)
and analyze more quantitatively their parity doublet and chiral
multiplet structure (see Fig. 1).

{\it \textbf{J}}\textbf{=1/2}: $N^+(2100)-N^-(2090)$;
$<PD_N>=2095$ MeV, $\delta_N=10$ MeV

\hskip 1.6cm $\Delta^+(1910)-\Delta^-(1900)$; $<PD_\Delta>=1905$
MeV, $\delta_\Delta=10$ MeV

\hskip 0.5cm Center of Gravity (c.g.)=2000 MeV; full mass spread
$=\delta=200$ MeV.

As we see, both $N$ and $\Delta$ doublets fit excellently the
definition of the parity doublets, but $N$ and $\Delta$ parity
doublets have very small correlation between them.

{\it \textbf{J}}\textbf{=3/2}: $N^+(1900)-N^-(2080)$;
$<PD_N>=1990$ MeV, $\delta_N=180$ MeV

\hskip 1.6cm $\Delta^+(1920)-\Delta^-(1940)$; $<PD_\Delta>=1930$
MeV, $\delta_\Delta=20$ MeV

\hskip 1.6cm c.g.=1960 MeV; $\delta=180$ MeV.

$\Delta$ doublet is a good parity doublet, but nucleon's doublet
hardly fits the parity doublet definition. It's interesting that
$\Delta$ parity doublet is located inside of the nucleon's parity
doublet. Another peculiar feature is that $\textbf{J}=3/2$ c.g.
is lower than $\textbf{J}=1/2$ c.g.

{\it \textbf{J}}\textbf{=5/2}: $N^+(2000)-N^-(2200)$;
$<PD_N>=2100$ MeV, $\delta_N=200$ MeV

\hskip 1.6cm $\Delta^+(1905)-\Delta^-(1930)$; $<PD_\Delta>=1918$
MeV, $\delta_\Delta=25$ MeV

\hskip 1.6cm c.g.=2009 MeV; $\delta=295$ MeV.

$\Delta$ doublet is a good PD, but nucleon doublet doesn't fit PD
definition. $N$ and $\Delta$ doublets are widely separated, and
have no overlap. Total spread is 295 MeV, and because of low $J$
we think this is the worst candidate for chiral multiplet.

{\it \textbf{J}}\textbf{=7/2} : $N^+(1990)-N^-(2190)$;
$<PD_N>=2090$ MeV, $\delta_N=200$ MeV

\hskip 1.6cm $\Delta^+(1950)-\Delta^-(2200)$; $<PD_\Delta>=2075$
MeV, $\delta_\Delta=250$ MeV

\hskip 1.6cm c.g.=2083 MeV; $\delta=250$ MeV.

Both $N$ and $\Delta$ doublets are too wide to qualify for PD's,
but their locations are correlated, with $N$ doublet lying inside
of the $\Delta$ doublet, making them a good chiral multiplet (see
Fig.1).

{\it \textbf{J}}\textbf{=9/2}: $N^+(2220)-N^-(2250)$;
$<PD_N>=2235$ MeV, $\delta_N=30$ MeV

\hskip 1.6cm $\Delta^+(2300)-\Delta^-(2400)$; $<PD_\Delta>=2350$
MeV, $\delta_\Delta=100$ MeV

\hskip 1.6cm c.g.=2293 MeV; $\delta=180$ MeV.

Both $N$ and $\Delta$ doublets have a fair chance to be considered
as a PD. But their relative position is quite dispersed.

{\it \textbf{J}}\textbf{=11/2, 13/2}, and \textbf{15/2}: there are
only two, two and one state here correspondingly, so there is no
point of discussing chiral multiplets for this sector.

Total average width for nucleons is $<\delta_N>=124$ MeV.

Total average width for deltas is $<\delta_\Delta>=81$ MeV.

As we se the deltas are much better PD's than the nucleons.

Grand average width for CG chiral multiplets
$<\delta_{GRAND}>=221$ MeV.

Original CG has 5 filled quartets. Quartets with {\it
\textbf{J}}\textbf{=1/2}, $\textbf{5/2}$, and $\textbf{9/2}$ have
no overlap between $N$ and $\Delta$ PD's, and only quartets with
{\it \textbf{J}}\textbf{=3/2}, $\textbf{7/2}$ could be considered
as a chiral multiplets.

Let's return again to the Table 4 and quantify the
Cohen-Glozman-Inopin multiplets (see Fig.1, right column).

{\it \textbf{J}}\textbf{=1/2}: $N^{+*}(1986)-N^{*-}(1897)$;
$<PD_N>=1942$ MeV, $\delta_N=89$ MeV

\hskip 1.6cm $\Delta^+(1910)-\Delta^-(1900)$; $<PD_\Delta>=1905$
MeV, $\delta_\Delta=10$ MeV

\hskip 1.6cm c.g.=1923 MeV; $\delta=89$ MeV.

Both $N$ and $\Delta$ doublets qualify as PD's. Delta doublet is
inscribed exactly inside the nucleon PD, making together a perfect
quartet (see Fig.1).

{\it \textbf{J}}\textbf{=3/2}: $N^+(1900)-N^{*-}(1895)$;
$<PD_N>=1898$ MeV, $\delta_N=5$ MeV

\hskip 1.6cm $\Delta^+(1920)-\Delta^-(1940)$; $<PD_\Delta>=1930$
MeV, $\delta_\Delta=20$ MeV

\hskip 1.6cm c.g.=1914 MeV; $\delta=45$ MeV.

Both $N$ and $\Delta$ doublets qualify as a PD's. $N$ and $\Delta$
doublets are clearly separated, but located close to each other,
with full width of only 45 MeV. This is the {\it best chiral
quartet} so far.

{\it \textbf{J}}\textbf{=5/2},$\textbf{7/2}$ and $\textbf{9/2}$:
CG quartets remained intact.

{\it \textbf{J}}\textbf{=11/2}: This line was unfilled in CG
model, and now we have a full quartet.

\hskip 1.6cm $N^+(2390)-N^-(2600)$; $<PD_N>=2495$ MeV,
$\delta_N=210$ MeV

\hskip 1.6cm $\Delta^+(2420)-\Delta^{*-}(2850)$;
$<PD_\Delta>=2635$ MeV, $\delta_\Delta=430$ MeV

\hskip 1.6cm c.g.=2565 MeV; $\delta=460$ MeV.

We see that both $N$ and $\Delta$ doublets have quite a large mass
spread, and hardly could qualify for a PD's. On the other hand
$N$ and $\Delta$ doublets overlap substantially, showing a clear
tendency for the chiral quartet creation. Although we have a big
total width for the {\it \textbf{J}}\textbf{=11/2} quartet, we
have to remember, that for such a high-momenta, baryon resonances
are becoming ill-defined (masses) and large spacings are natural.

{\it \textbf{J}}\textbf{=13/2}: CG has two states. Now we have:

\hskip 1.6cm $N^+(2700)-N^-(2900)$; $<PD_N>=2800$ MeV,
$\delta_N=200$ MeV

\hskip 1.6cm $\Delta^{*+}(3200)-\Delta^-(2750)$;
$<PD_\Delta>=2975$ MeV, $\delta_\Delta=450$ MeV

\hskip 1.6cm c.g.=2888 MeV; $\delta=500$ MeV.

As we see, both $N$ and delta doublets have a very large mass
spread, and cannot qualify as a PD. There is clear overlap
between $N$ and $\Delta$ doublets, which indicates attraction and
possible chiral multiplet creation (after the masses will be
better measured).

{\it \textbf{J}}\textbf{=15/2}: CG has only one state here. Now
we have:

\hskip 1.6cm $N^+(3000)-N^{*-}(3100)$; $<PD_N>=3050$ MeV,
$\delta_N=100$ MeV

\hskip 1.6cm $\Delta^+(2950)-\Delta^-(3480)$; $<PD_\Delta>=3215$
MeV, $\delta_\Delta=530$ MeV

\hskip 1.6cm c.g.=3133 MeV; $\delta=530$ MeV.

Nucleon doublet definitely fit as a PD. Delta doublet is no
good as a PD. It is wonderful that $N$ doublet is inscribed
exactly inside $\Delta$ doublet, making this quartet a chiral
multiplet (see Fig.1).

{\it \textbf{J}}\textbf{=17/2}: CG has zero states in this line.
Now we have:

\hskip 1.6cm $N^{*+}(3500)-N^-(3450)$; $<PD_N>=3475$ MeV,
$\delta_N=50$ MeV

\hskip 1.6cm $\Delta^+(3490)-\Delta^{*-}(3300)$;
$<PD_\Delta>=3395$ MeV, $\delta_\Delta=190$ MeV

\hskip 1.6cm c.g.=3435 MeV; $\delta=200$ MeV.

Nucleon doublet is happened to be a perfect PD with mass spread
of only 50 MeV. Delta doublet is a fair candidate for the PD. Two
doublets have almost 100\% overlap, making a real chiral multiplet
with total spread of only 200 MeV for such a high momentum (see
Fig.1).

{\it \textbf{J}}\textbf{=19/2}: CG has zero states in this line.
Now we have a quartet. Nucleon doublet is 250 MeV wide, and
$\Delta$ doublet is 500 MeV wide. New data and PWA could
considerably shrink this quartet. $N$ doublet is inscribed
exactly inside $\Delta$ doublet.

\hskip 2.5cm $N^{+}(3500)-N^{*-}(3750)$; $<PD_N>=3625$ MeV,
$\delta_N=250$ MeV

\hskip 2.5cm $\Delta^{*+}(3500)-\Delta^-(4000)$;
$<PD_\Delta>=3750$ MeV, $\delta_\Delta=500$ MeV

\hskip 3.5cm c.g. $=3688$ MeV; $\delta=500$ MeV

{\it \textbf{J}}\textbf{=21/2}: CG has zero states in this line.
So far we've got single $N^-$ and single $\Delta^-$, lying close
to each other. Hopefully soon $N^+$ and $\Delta^+$ partners will
be discovered to form a chiral multiplet.

\hskip 3.5cm $N^-(3950)-\Delta^{*-}(4100)$;
$\delta=150$ MeV
\vskip 0.3cm
Finally we compare full mass spreads
for original CG (Table 1) and modified CGI schemes (Table 4). For
the totally filled quartets {\it \textbf{J}}\textbf{=1/2,.....1/9}
we have: \vskip 0.3cm
\hskip 2.5cm $<\delta_{grand}>_{CG}=221$
MeV ; $\quad <\delta_{grand}>_{CGI}=172$ MeV

\vskip 0.3cm

Clearly, in the modified CGI scheme we have much {\it
more narrow} quartets, making the conjecture of chiral multiplets
plausible. If we consider CGI multiplets for {\it
\textbf{J}}\textbf{=11/2,.....21/2}, where CG scheme failed, we
get dispersed mass spreads from $\delta=150$ MeV to $\delta=530$
MeV, with $<\delta_{grand}>=390$ MeV.

We have to admit that these results contradicts the basic CG idea:
chiral symmetry is restored after 2 GeV mass region, and one
should see clear chiral multiplets, which has to be {\it more
degenerate} in high $J$, $M$ region. In fact, one can see
multiplet's spread growing with $J$, $M$. Klempt came recently to
the similar conclusions from a different venue \cite{8}.

\vskip 1.0cm

{\bf \ \ \ \ \ \ \ \ \ \ \ \ \ \ \ \ \ \ \ \ \ \ \ \ \ \ \ \ \ \ \
\ \ \ \ b) \underline {Klempt Model}}

\vskip 0.5cm

Recently Klempt published a series of papers devoted to baryon
spectroscopy \cite{8}.

Based on analysis of data, Klempt proposed a new baryon mass
formula:

\begin{equation}
M^2=M^2_\Delta+{{n_sM^2_s}\over{3}}+a(L+N)-s_iI_{sym},
\label{1}
\end{equation}
where $M^2_s=(M^2_\Omega-M^2_\Delta)$, $s_i=(M^2_\Delta-M^2_N)$,
$n_s$ the number of strange quarks in a baryon, $L$ the intrinsic
orbital angular momentum. $N$ is the principal quantum number.
$I_{sym}$ is the fraction of the wave function (normalized to the
nucleon wave function) which is antisymmetric in spin and flavor.

The author claimed that formula (\ref{1}) reproduces nearly all
known baryon masses. But it is evident that Eq.(\ref{1}) is
based on the presumption that any baryon state clearly can be
defined as a {\it pure state} in $L$, $N$, $I_{sym}$ functional
space. In other words,there is no admixtures with different "$L$,
$N$ - orbitals" in a given baryon WF. But we know from our
experience and the results of many classical papers, that at least
for some of the BR's this could be 100\% wrong. Just take for
example $N_{{3/2}^+}(1720)$ B.R. It's defined as a strong
superposition of five $L$-orbitals, with its major component
having only about 40-50\% of all wave function.

We will analyze Klempt's findings on clustering and chiral
properties of baryon sector.

The most interesting are the multiplets with assigned $S=3/2$
intrinsic spin. First he identified the "stretched" states with
$J=L+S$; $L=0,1....,6$ and $S=3/2$, i.e. resonances with quantum
numbers $$ J^P=3/2^+, 5/2^-, 7/2^+, 9/2^-, 11/2^+, 13/2^-,
15/2^+$$

These resonances are shown in Table 5 in the fifth column. Omitted
are the decuplet ground states $(L=0)$ which also fall into this
category. In the nonrelativistic quark model (NRQM), we expect
{\it single} resonances for $L=0$ (the ground states), {\it
triplets} for $L=1$, and {\it quartets} for higher $L$. The
multiplet structure is clearly visible in Table 5, even though the
multiplets are not complete.

As we look more carefully at this scheme in Table 5, we uncover an
extremely interesting link between the dynamics and symmetry. When
we construct the tower of states, the first floor with $L=1$ is:

\hskip 5.7 cm {\it \textbf{L}}\textbf{=1}: $N\rightarrow \Sigma
\rightarrow \Lambda \rightarrow \Delta$

\newpage
The second floor is:

\hskip 5.7 cm {\it \textbf{L}}\textbf{=2}: $\Delta \rightarrow
\Sigma \rightarrow \Lambda \rightarrow N$

The third floor is:

\hskip 6.5 cm {\it \textbf{L}}\textbf{=3}: $ N \rightarrow \Delta$

The fourth floor is:

\hskip 7.1 cm {\it \textbf{L}}\textbf{=4} :$ \Delta,$

where by $N,\Delta,\Sigma,\Lambda$ here we understand
corresponding multiplets of $N^*,\Delta^*,\Sigma^*,\Lambda^*$,
from the Table 5 (see Fig.2).

There is evidently a new symmetry group that arises, which
reflects the total isospin degeneracy for a given floor:

$$(I=0)+(I=1/2)+(I=1)+(I=3/2)\rightarrow Supermultiplet$$It seems
that at low $L=1$ we have a grand $SU_I(4)$ group, for
the $L=2$ it is still $SU_I(4)$, then for $L=3$ it decays into
{\it chain of the subgroups}, represented by ${SU_I(2)}$, then for
$L=4$ \hskip 0.5 cm $SU_I(2)$ subsequently decays into $SU_I(1)$.
(We omit here singlets with $L=5,6$.) We can think of the
following chain:

$$SU^1_I(4) \rightarrow SU^2_I(4) \rightarrow SU_I(2) \rightarrow SU_I(1)$$


\begin{table}[ht]
\caption{\textbf{Klempt's $\textbf{S=3/2}$ Multiplets}}
\begin{center}
\begin{tabular}{| c | c | c | c | c | c |}
\hline
{\it \textbf{L}} & {\it \textbf{J}}{\it \textbf{=L}} \textbf{-3/2} & {\it \textbf{J}}{\it
\textbf{=L}} \textbf{-1/2} & {\it \textbf{J}}{\it \textbf{=L}} \textbf{+1/2}& {\it
\textbf{J}}{\it \textbf{=L}} \textbf{+3/2} & \textbf{Average}\\
\hline
1 & - & $N_{{1/2}^-}(1650)$ & $N_{{3/2}^-}(1700)$ & $N_{{5/2}^-}(1675)$ & 1675 \\
\hline
1 & - & $\Sigma_{{1/2}^-}(1750)$ & - & $\Sigma_{{5/2}^-}(1675)$ & 1763 \\
\hline
1 & - & $\Lambda_{{1/2}^-}(1800)$ & - & $\Lambda_{{5/2}^-}(1830)$ & 1815 \\
\hline
1 & - & $\Delta_{{1/2}^-}(1900)$ & $\Delta_{{3/2}^-}(1940)$ & $\Delta_{{5/2}^-}(1930)$ &
1923 \\
\hline
2 & $\Delta_{{1/2}^+}(1910)$ & $\Delta_{{3/2}^+}(1920)$ & $\Delta_{{5/2}^+}(1905)$ &
$\Delta_{{7/2}^+}(1950)$ & 1921 \\
\hline
2 & - & $N_{{3/2}^+}(1900)$ & $N_{{5/2}^+}(2000)$ & $N_{{7/2}^+}(1990)$ & 1963 \\
\hline
2 & - & $\Sigma_{{3/2}^+}(2080)$ & $\Sigma_{{5/2}^+}(2070)$ & $\Sigma_{{7/2}^+}(2030)$ &
2060 \\
\hline
2 & - & - & $\Lambda_{{5/2}^+}(2110)$ & $\Lambda_{{7/2}^+}(2020)$ & 2065 \\
\hline
3 & - & $N_{{5/2}^-}(2200)$ & $N_{{7/2}^-}(2190)$ & $N_{{9/2}^-}(2250)$ & 2213 \\
\hline
3 & - & $\Delta_{{5/2}^-}(2350)$ & - & $\Delta_{{9/2}^-}(2400)$ & 2375 \\
\hline
4 & - & $\Delta_{{7/2}^+}(2390)$ & $\Delta_{{9/2}^+}(2300)$ & $\Delta_{{11/2}^+}(2420)$ &
2370 \\
\hline
5 & - & - & - & $\Delta_{{13/2}^-}(2750)$ & 2750 \\
\hline
6 & - & - & - & $\Delta_{{15/2}^+}(2950)$ & 2950 \\
\hline
\end{tabular}
\end{center}
\end{table}

As one can see from Fig.2 we significantly generalize Klempt's
model and create a supermultiplets in the "Isospin Space" instead
of Klempt's "L-Space" multiplets. New supermultiplets are
represented in Fig.2 as a creature with four floors. As we see,
the topology realized as plane $(D=2) \rightarrow$ plane $(D=2)
\rightarrow$ line $(D=1) \rightarrow$ point $(D=0)$. Our own
analysis of the PDG2002 data lead to the different supermultiplets
shown in a Figure 3. We will return to the comparison of
different models a bit later.

An interesting feature of the Klempt's $S=3/2$ multiplets is a
"good definition" of a floor structure, i.e. different floors are
really separated in the mass. The "widths" of the floors with
increasing $L$ are (in MeV): $248 (L=1) \rightarrow 144(L=2)
\rightarrow 162(L=3) \rightarrow 0(L=4)$. So they are smoothly
decreasing with $L$ from 248 MeV to zero at $L=4$ (see Fig. 4).

One can see, that beginning from the $L=2$ the {\it quartet
structure} is allowed and arise in reality. This is reminiscent
of the CG model where the $N$,$\Delta$-quartets appeared from the
chiral symmetry.

Let's make a detailed analysis of the multiplet's mass spreads
from the Table 5. \vskip 0.5 cm \hskip 4.5 cm {\it \textbf{L}}
\textbf{=1}: $N(1650 - 1700), \delta = 50 MeV$
$$\hskip 1.2 cm \Sigma (1750 - 1775), \delta = 25 MeV$$
$$\hskip 1.2 cm \Lambda(1800 - 1830), \delta = 30 MeV$$
$$\hskip 1.2 cm \Delta(1900 - 1940), \delta = 40 MeV$$For the $L=1$
floor individual mass spreads vary from 25 to 50
MeV, and $<\delta_{L=1}>=36$ MeV only. Total width of the L=1
supermultiplet is 290 MeV, with c.g.$=1794$ MeV.

\vskip 0.5 cm
\hskip 4.5 cm {\it \textbf{L}} \textbf{=2}:
$\Delta(1905 - 1950), \delta = 45 MeV$
$$\hskip 1.3 cm N (1900 - 2000), \delta = 100 MeV$$
$$\hskip 1.2 cm \Sigma(2030 - 2080), \delta = 50 MeV$$
$$\hskip 1.2 cm \Lambda(2020 - 2110), \delta = 90 MeV$$For the $L=2$
floor individual mass spreads vary from 45 to 100
MeV, and $<\delta_{L=2}>=71$ MeV. Total width of $L=2$
supermultiplet is 210 MeV, with c.g. $= 2002$ MeV. One can see
that total width of the $L=2$ floor is {\it diminished}, compared
to the $L=1$ floor.

\vskip 0.5 cm \hskip 4.5 cm {\it \textbf{L}} \textbf{=3}: $N(2190
- 2250), \delta = 60 MeV$
$$\hskip 1.2 cm \Delta(2350 - 2400),\delta = 50 MeV$$For the $L=3$
floor individual mass spreads vary from 50 to 60
MeV, with $<\delta_{L=3}>=55$ MeV. Total width of the $L=3$ floor
is 210 MeV, the same as $L=2$, with c.g. $= 2294$ MeV.

\vskip 0.5 cm
\hskip 4.5 cm {\it \textbf{L}} \textbf{=4}: $\Delta(2300 - 2420),
\delta = 120 MeV$ \vskip 0.3 cm $L=4$ supermultiplet
shrinks down to the $\Delta$ triplet, but it's might be just
experimental problem which will be resolved in coming years. It's
evident that our supermultiplets are getting {\it narrower} with
$L$ increasing and the masses increasing, which roughly resembles
the chiral symmetry restoration scenario.

In his last paper in the series \cite{8}, Klempt discussed the
mass spectrum of $N$, $\Delta$ resonances only. The author
compared CG conjecture with the possibility that {\it high-mass}
states are organized into $(L,S)$-multiplets with defined
intrinsic quark spins and orbital angular momenta. Table 6, which
is adapted from the CG papers, shows $N^*$ and $\Delta^*$ masses
above 1.9 GeV, for states with positive and negative parity. In
many cases, the effect of PD is {\it striking}; states with
identical {\it \textbf{J}} but opposite parity often have very
similar masses. This does not of course automatically imply that
parity doublets are generated by restoration of chiral symmetry.
Consider the first six $\Delta$ states in Table 6 with {\it
\textbf{J}}$=1/2$,
$3/2$, and $5/2$. The masses are clearly degenerate; they
form three PD's.

\begin{table}[ht]
\caption{\textbf{CG versus Klempt Models}}
\begin{center}
\begin{tabular}{| c | l | c | l | c |}
\hline
\large{{\it \textbf{J}}$=1/2$} & \large{ {1}$ \hskip 0.2 cm
N_{{1/2}^+}(2100)$} &\large{$N_{{1/2}^-}(2090)$}
&\large{{a} $\hskip 0.2 cm \Delta_{{1/2}^+}(1910)$} &
\large{$\Delta_{{1/2}^-}(1900)$} \\
\hline
\large{{\it \textbf{J}}$=3/2$} & \large{{2}$ \hskip 0.2 cm
N_{{3/2}^+}(1900)$} &\large{$N_{{3/2}^-}(2080)$} &
\large{{b} $\hskip 0.2 cm \Delta_{{3/2}^+}(1920)$} &
\large{$\Delta_{{3/2}^-}(1940)$} \\
\hline
\large{{\it \textbf{J}}$=5/2$} & \large{{3}$\hskip 0.2 cm
N_{{5/2}^+}(2000)$} &\large{$N_{{5/2}^-}(2200)$} &
\large{{c} $\hskip 0.2 cm \Delta_{{5/2}^+}(1905)$} &
\large{$\Delta_{{5/2}^-}(1930)$} \\
\hline
\large{{\it \textbf{J}}$=7/2$} & \large{{4}$\hskip 0.2 cm
N_{{7/2}^+}(1990)$} &\large{$N_{{7/2}^-}(2190)$} &
\large{{d} $\hskip 0.2 cm \Delta_{{7/2}^+}(1950)$} &
\large{$\Delta_{{7/2}^-}(2200)$} \\
\hline
\large{{\it \textbf{J}}$=9/2$} & \large{{5}$\hskip 0.2 cm
N_{{9/2}^+}(1990)$} &\large{$N_{{9/2}^-}(2250)$} &
\large{{e} $\hskip 0.2 cm \Delta_{{9/2}^+}(2300)$}&
\large{$\Delta_{{9/2}^-}(2400)$} \\
\hline
\large{{\it \textbf{J}}$=11/2$} &
\large{ {6} {$\hskip 0.2 cm N_{{11/2}^+}$}}
&\large{$N_{{11/2}^-}(2600)$} & \large{ {f} $\hskip 0.2 cm
\Delta_{{11/2}^+}(2420)$} & \large{{$\Delta_{{11/2}^-}(*)$}} \\
\hline
\large{{\it \textbf{J}}$=13/2$} & \large{{7}$\hskip 0.2 cm
N_{{13/2}^+}(2700)$} &\large{{$N_{{13/2}^-}$}} &
\large{ {g}  {$\hskip 0.2 cm
\Delta_{{13/2}^+}$}} & \large{$\Delta_{{13/2}^-}(2750)$} \\
\hline
\large{{\it \textbf{J}}$=15/2$} &
\large{{8} {$\hskip 0.2 cm
N_{{15/2}^+}$}}&\large{ {$N_{{15/2}^-}$}} &
\large{{h} $\hskip 0.2 cm \Delta_{{15/2}^+}(2950)$} &
\large{ {$\Delta_{{15/2}^-}(*)$}} \\
\hline
\end{tabular}
\end{center}
\end{table}
(Klempt suggested that the states marked with a $(*)$ in Table 6
should have considerably higher masses than their chiral partners
while the other five states in color should be degenerate in mass
with corresponding states of opposite parity).

The $\Delta_{{7/2}^+}(1950)$ and the $\Delta_{{7/2}^-}(2200)$
should also form a PD but the $\Delta_{{7/2}^+}(1950)$ has a mass
which is very close to the other three positive-parity BR. These
four positive-parity $\Delta$'s rather seem to belong to a {\it
spin quartet} of states with $L=2$ and $S=3/2$ coupling to {\it
\textbf{J}}$=1/2,...7/2$. The question arises if the PD's are
really due to restoration of chiral symmetry or do the PD's
reflect a symmetry of the underlying quark dynamics?

The CG model requires the existence of a $\Delta_{{11/2}^-}$ and
$N_{{11/2}^+}$ at about 2500 MeV, of a $\Delta_{{13/2}^+}$ and a
$N_{{13/2}^+}$ at 2750 MeV, and of three additional states at 2950
MeV. Klempt investigated if the occurrence of PD's can be
understood naturally within SU(6) multiplet structure of BR. In
this interpretation PD's occur naturally, but the prediction for
so-far unobserved PD's differs.

A decision, if $N^*$'s form PD's, requires a quantitative
analysis. First we notice that according to Eq.(\ref{1}) the mass
difference between two resonances with consecutive $L$ and
otherwise identical quantum numbers vanishes asymptotically:

\begin{equation}
M^2_{L+1}-M^2_L=(M_{L+1}-M_L)(M_{L+1}+M_L)=a,
\label{2}
\end{equation}

hence $$M_{L+1}-M_L={{a}\over{(M_{L+1}+M_L)}}$$

\vskip 0.5 cm
Asymptotically, all mass separations vanish with $1/M$ and chiral
symmetry is trivially restored. Klempt finally compared the
consistency of the data with the assumption of parity doublets
and, alternatively, with their consistency with $(L,S)$ multiplets
with vanishing {\it \textbf{LS}} coupling. But the author didn't
in fact compare the consistency of the chiral multiplets
assumption, which is of paramount importance.

First Klempt computed the mean mass deviation of BR's when they
are interpreted as parity doublets:

\begin{equation}
\sigma_{PD's}=\sqrt{{{1}\over{10}}\sum_{i=1,20}(M_i-M_{\pm})^2}=97 MeV
\label{3}
\end{equation}
where $M_{\pm}$ are the mean masses of positive- and
negative-parity resonances {\it paired} to one parity doublet (see
Table 6). Now author determined the deviation of baryon masses
from the mean value of a $(L,S)$ multiplet:

\begin{equation}
\sigma_{spin \hskip 0.1 cm multiplets}=\sqrt{{{1}\over{13}}\sum_{i=1,20}(M_i-M_{cg})^2}=39 MeV
\label{4}
\end{equation}
where the $M_{cg}$ are the mean values (centers of gravity). The
comparison of the two hypotheses reveals that evidence for PD's in
the high-mass spectrum is {\it weak}, at most. The data is better
described in terms of $(L,S)$-multiplets embracing $SU(6)$
multiplets of different $J$ but having the same $L$ and $S$.
Klempt claimed that the symmetry leading to PD's is the vanishing
of spin-orbit forces and not a {\it phase transition} to chiral
dynamics. We have to note that CG a few times stressed that this
mechanism is rather a {\it crossover} and not a phase transition
\cite{4}.

As we have seen in the previous section, the failure of the CG
scheme on PD's and chiral multiplets also follows from the careful
examining of PD's and chiral multiplet's widths and spacings with
growing $J$ and $M$.

\vskip 0.7 cm

{\bf \ \ \ \ \ \ \ \ \ \ \ \ \ \ \ \ \ \ \ \ \ \ \ \ \ \ \ \ \ \ \
\ \ \ \ c) \underline {Data Analysis}}

\vskip 0.4 cm

{\bf \ \ \ \ \ \ \ \ \ \ \ \ \ \ \ \ \ \ \ \ \ \ \ \ \ \ \ \ \ \ \
\ \ \ \ \ \ \ \ \ \ \ \ {\Large{$N-\Delta$}}} \vskip 0.4 cm

Clustering structure of the baryonic spectra was revealed recently
by us by examining the PDG2000 issue \cite{9},\cite{10} (there is
no changes in baryonic sector in PDG2002). Let us briefly remind
the specifics of this analysis. The nonstrange sector is very
rich, comprising 23 N and 22 states. We will include in the
analysis three new resonances, recently discovered at ELSA,
SAPHIR: $N_{{3/2}^-}(1895)$, $N_{{1/2}^-}(1897)$, and
$N_{{1/2}^+}(1986)$. We will also include in the analysis the
so-called $N$($\sim 3000$ Region) and $\Delta$($\sim 3000$ Region)
(see Table 2). So, altogether we have 31 $N$ and 28 $\Delta$
states. $N$ and $\Delta$ spectra exhibit very interesting
clustering properties. In the nucleon sector we see the following
four clusters: $sextet$, $\Delta=70 MeV$:

$$N_{{1/2}^-}(1650)-N_{{5/2}^-}(1675)-N_{{5/2}^+}(1680)-N_{{3/2}^-}(1700)-N_{{1/2}^+}(1710)-N_{{3/2}^+}(1720)$$

\vskip 0.5 cm

{\it triplet}: \hskip 0.8 cm
$N_{{3/2}^-}(1895)-N_{{1/2}^-}(1897)-N_{{3/2}^+}(1900), \delta=5
MeV$

\vskip 0.7 cm

{\it triplet}: \hskip 0.8 cm
$N_{{3/2}^-}(2080)-N_{{1/2}^-}(2090)-N_{{1/2}^+}(2100), \delta=20
MeV$

\vskip 0.7 cm

{\it quartet}: \hskip 0.6cm
$N_{{7/2}^-}(2190)-N_{{5/2}^-}(2200)-N_{{9/2}^+}(2220)-N_{{9/2}^-}(2250),
\delta=60 MeV$

\vskip 0.7 cm

So the average width is $39 MeV$ only. First $N$-cluster is split
into three PD's: \vskip 0.4 cm \hskip 3 cm
$N_{{1/2}^-}(1650)-N_{{1/2}^+}(1710),\delta=60 MeV$ \vskip 0.2 cm
\hskip 3 cm $N_{{3/2}^-}(1700)-N_{{3/2}^+}(1720),\delta=20 MeV$
\vskip 0.2 cm \hskip 3 cm
$N_{{5/2}^-}(1675)-N_{{5/2}^+}(1680),\delta=5 MeV$ \vskip 0.2 cm

Second $N$-cluster has one parity doublet:
\vskip 0.4 cm \hskip 3 cm
$N_{{3/2}^-}(1895)-N_{{3/2}^+}(1900), \delta=5 MeV$ \vskip 0.2 cm

Third $N$-cluster has one parity doublet:
\vskip 0.4 cm \hskip 3 cm
$N_{{1/2}^-}(2090)-N_{{1/2}^+}(2100), \delta=10 MeV$
\vskip 0.2 cm
Fourth $N$-cluster has one parity doublet:

\vskip 0.4 cm \hskip 3cm
$N_{{9/2}^+}(2220)-N_{{9/2}^-}(2250), \delta=30 MeV$ (see Figs. 5-6)
\vskip 0.7 cm

One can see that our PD's structure of $N$-spectrum {\it differs
drastically} from the CG picture: most of the parity doublets
appears in low energy region, and the rest of them are sparsely
distributed over the high energy region. Our PD's also much
better fit the definition: mass spreads vary from $5$ to $60$ MeV,
with $<\delta_N>=22$ MeV only.

In $\Delta$ sector we find the following two clusters:

{\it septet:}

$$\Delta_{{1/2}^-}(1900)-\Delta_{{5/2}^+}(1905)-\Delta_{{1/2}^+}(1910)-\Delta_{{3/2}^+}(1920)-$$
$$ -\Delta_{{5/2}^-}(1930)-\Delta_{{3/2}^-}(1940)-\Delta_{{7/2}^+}(1950), \Delta=50
MeV$$

{\it triplet:}
$$\Delta_{{7/2}^+}(2390)-\Delta_{{9/2}^-}(2400)-\Delta_{{11/2}^+}(2420),\delta=30 MeV$$

{\it First} $\Delta$-cluster is split into three PD's plus one
extra state:
$$\Delta_{{1/2}^-}(1900)-\Delta_{{1/2}^+}(1910), \delta=10 MeV$$
$$\Delta_{{3/2}^+}(1920)-\Delta_{{3/2}^-}(1940), \delta=20 MeV$$
$$\Delta_{{5/2}^+}(1905)-\Delta_{{5/2}^-}(1930), \delta=25 MeV$$

{\it Second} cluster has no PD's.

One can see that $\Delta$-clusters are shifted upwards as a whole
against $N$ clusters, playing more in accord with CG conjecture.
$N$-cluster's center of gravity lying at $1883$ MeV, and
$\Delta$-cluster's c.g lies at $2067$ MeV, resulting in $184$ MeV
difference (see Figs. 5, 6).

Let's look at our picture of parity doublets, whether we can
create some chiral multiplets. It is quite evident that only
sector with $J^P=3/2^{+-}$ gives a good sample of chiral
multiplet:

\vskip 0.5 cm \hskip 2 cm {\it \textbf{J}}\textbf{=3/2} :
$(N^+(1900), N^-(1895), \Delta^+(1920), \Delta^-(1940)),
\delta=45 MeV$

\vskip 0.6 cm
{\bf \ \ \ \ \ \ \ \ \ \ \ \ \ \ \ \ \ \ \ \ \ \ \ \ \ \ \ \ \ \ \
\ \ \ \ \ \ \ \ \ {\Large{$\Lambda-\Sigma$}}} \vskip 0.4 cm
Because $\Lambda-\Sigma$ BR's have only one strange quark,
their shape is still not so deformed. If we can imagine that we
have 3 balls in a bag, and two of them are of almost the same
weight, while third a bit heavier than the two, the bag will take
the form of a pear. It will be reflectionally asymmetric. {\it
Near the rest}, the deformation is perhaps still not so dramatic
and the lowest excitations are similar to those of the nonstrange
baryons. If the heavier ball starts to gain rotational energy,
the deformation will increase. The pear shape gets more pronounced
and when the pear {\it oscillates} it gives rise to parity
doublets, which we already see.

Full listing \cite{5} give to us $18 \quad \Lambda$ and $26
\quad \Sigma$ BR's. Some of the states are lacking the $J^P$
assignments. State $\Lambda(2000)$ does not have $J^P$, but data
from Cameron 78 allowed tentatively, the $J^P=1/2^-$ assignment.
Further evidence came from the recent paper by Iachello \cite{11}
and older one by Capstick-Isgur \cite{12}. Therefore we assign
$J^P=1/2^-$ to the $\Lambda(2000)$. The $\Lambda$-states with
highest masses, $\Lambda(2350)$ and $\Lambda(2585)$ were not
described theoretically and there are no clear claims from the
experiments. For this reason we will not include $\Lambda(2350)$,
$\Lambda(2585)$ in our analysis.

The situation with $\Sigma$ hyperons is even more interesting. Two
low-lying $\Sigma(1480)$ and $\Sigma(1560)$ do not have any $J^P$
assignments from the experiment and theory can't predict them
either. We will exclude $\Sigma(1480)$, $\Sigma(1560)$ from our
analysis. The production experiments \cite{5} give strong evidence
for $\Sigma(1620)$, tentatively claiming $J^P=1/2^+$. This claim
is in accord with Iachello \cite{11}. So with newly defined
$\Sigma^*_{{1/2}^+}(1620)$, we form an {\it exact} parity doublet
$\Sigma_{{1/2}^-}(1620)-\Sigma^*_{{1/2}^+}(1620)$. The production
experiments \cite{5} give strong evidence for $\Sigma(1670)$ bumps
without $J^P$ assignments. Using predictions by Iachello and
Isgur, we clearly get $J^P=1/2^-$ for $\Sigma(1670)$. It's
interesting that this way we have two resonances with the same
mass and different $J^P$. The state \hskip 0.2 cm
$\Sigma(1690)$\hskip 0.2 cm has\hskip 0.2 cm most \hskip 0.2
cmlikely \hskip 0.2 cmclaim from the data as $J^P=5/2^+$. We will
assign $J^P=5/2^+$ to $\Sigma(1690)$ in our analysis. Next
$\Sigma$ without $J^P$ assignment will be $\Sigma(2250)$. Using
results from Iachello and Isgur we assigned $J^P=5/2^-$ to
$\Sigma(2250)$. Last few bumps, $\Sigma(2455)$, $\Sigma(2620)$,
$\Sigma(3000)$, and $\Sigma(3170)$ has no experimental claims for
$J^P$, and there are no theoretical predictions so far for such a
high masses. For this reason we will not include $\Sigma(2455)$,
$\Sigma(2620)$, $\Sigma(3000)$, $\Sigma(3170)$ in our analysis.

Clustering pattern is very nontrivial in $\Sigma$ spectrum. We
clearly see four clusters here.

{\it Doublet:} \hskip 1 cm $\Sigma_{{1/2}^-}(1620)-
\Sigma^*_{{1/2}+}(1620)$

{\it Quartet:} \hskip 1 cm
$\Sigma_{{1/2}+}(1660)-\Sigma_{{3/2}^-}(1670)-\Sigma^*_{{1/2}-}(1670)-
\Sigma^*_{{5/2}^+}(1690),\delta=30 MeV$

{\it Triplet:} \hskip 1.2 cm
$\Sigma_{{1/2}^-}(1750)-\Sigma_{{1/2}^+}(1770)-\Sigma_{{5/2}^-}(1775),
\delta= 25 MeV$

{\it Triplet:} \hskip 1.2 cm
$\Sigma_{{5/2}^+}(2070)-\Sigma_{{3/2}^+}(2080)-\Sigma_{{7/2}^-}(2100),
\delta= 30 MeV$

One can see that our $\Sigma$-clusters are mostly grouped in low
energy region (9 states), and only one cluster (triplet) is
located above 2000 MeV. This result clearly contradicts basic CG
conjecture.

{\it First} cluster is just a parity doublet:

\hskip 5 cm $\Sigma_{{1/2}^-}(1620)-\Sigma^*_{{1/2}^+}(1620).$

{\it Second} cluster has one parity doublet:

\hskip 5 cm $\Sigma_{{1/2}^-}(1660)-\Sigma^*_{{1/2}^-}(1670).$

{\it Third} cluster has one parity doublet:

\hskip 5 cm $\Sigma_{{1/2}^-}(1750)-\Sigma_{{1/2}^+}(1770).$

{\it Fourth} cluster has no parity doublets. We still have to
understand why $\Sigma$-sector produces PD's with $J^P=1/2^{+-}$
only.

In $\Lambda$ sector we witness only one cluster; this is {\it
quartet}:
$$\Lambda_{{1/2}^-}(1800)-\Lambda_{{1/2}^+}(1810)-\Lambda_{{5/2}^+}(1820)-\Lambda_{{5/2}^-}(1830),
\delta=30 MeV$$
This quartet is split into two parity doublets:

$$\Lambda_{{1/2}^-}(1800)-\Lambda_{{1/2}^+}(1810),
\delta=10 MeV$$

$$\Lambda_{{5/2}^+}(1820)-\Lambda_{{5/2}^-}(1830),
\delta =10 MeV$$ $\Sigma$-cluster's $c.g.=1790$ MeV,
$\Lambda$-cluster's $c.g.=1815$ MeV, so there is a good overlap
between the $\Lambda-\Sigma$ clusters in a global sense (see Figs.
5-7).

From our $\Lambda-\Sigma$ clusters we can construct only one
chiral multiplet:

\vskip 0.2 cm \hskip 1.5 cm {\it \textbf{J}}\textbf{=1/2} :
$(\Sigma^+(1770), \Sigma^-(1750), \Lambda^+(1810),
\Lambda^-(1800)), \delta=60 MeV$ \vskip 0.2 cm One can see that
$\Lambda-\Sigma$ chiral multiplet is located in a low-energy
region, in contrast with CG predictions (see Figure 8 for overall
pattern in $N-\Delta-\Lambda-\Sigma$ sector).

On Figure 9 we have made a comparison between Klempt's $S=3/2$
multiplets and our clusters. For \hskip 0.2 cm the \hskip 0.2 cm
most, \hskip 0.2 cm the \hskip 0.2 cm two schemes have a very good
overlap. In particular our $1\Delta(1900-1950)$ split into two
Klempt's multiplet $2L=1(1900-1940)$ and $2L=2(1905-1950)$ \hskip
0.2 cmmaking 100\%\hskip 0.2 cm
overlap. Klempt's\hskip 0.2 cm multiplet
\hskip 0.2 cm $1L=4(2300-2420)$ \hskip 0.2 cm included \hskip 0.2
cmhis \hskip 0.2 cm$2L=3(2350-2400)$ and our $2\Delta$ multiplet,
making a {\it double} overlap. Few of our multiplets, do not have
corresponding partners: $3N$, $1\Sigma$ and $2\Sigma$.

\vskip 0.5 cm

{\bf \ \ \ \ \ \ \ \ \ \ \ \ \ \ \ \ \ \ \ \ \ \ \ \ \ \ \ \ \ \ \
\ \ \ \ d) \underline {Kirchbach Model}}

\vskip 0.2 cm

The history of spectroscopy is basically the history of finding
the degeneracy symmetry of the spectra. It is natural to expect
from the models of hadron structure that they should supply us
with a satisfactory description of baryon excitations. The
knowledge on the degeneracy group of baryon spectra appears as a
key tool in constructing the underlying strong interaction
dynamics. To uncover it, one has first to analyze isospin by
isospin how the masses of the resonances from the full listing in
PDG spread with spin and parity. Such an analysis has been
performed by Kirchbach \cite{13}, where it was found that
Breit-Wigner masses reveal on $M/J$ plane a well pronounced spin
and parity clustering. There, it was further shown, that the
quantum numbers of the resonances belonging to a particular
cluster fit into Lorentz group representations of the type
$\{k/2,k/2\}\otimes [\{1/2,0\}\oplus \{0,1/2\}]$, known as
Rarita-Schwinger (RS) fields. To be specific, one finds the {\it
three} RS clusters with $k = 1,3,$ and 5 in both $N$ and $\Delta$
spectra. These representations accommodate states with different
spins and parities and will be referred to as {\it
multispin-parity} clusters (MPC). To illustrate this statement it
is useful to recall that the irreducible representations (irreps)
$\{k/2, k/2\}$ of the Lorentz group yield in its Wick rotated
compact form $O(4)$, four-dimensional ultraspherical harmonics,
here denoted by $\sigma_\eta$, with $\sigma=k+1$, where $\sigma$
is related to the principal quantum number of the Coulomb problem,
while $\eta$ is related to parity. These $O(4)$ irreps will be
occasionally referred to in the following as Coulomb multiplets.
The latter are known to collect mass degenerate $O(3)$ states of
integer internal angular momenta, $L$, with $L=0,...,\sigma-1$.
All three-dimensional spherical tensors (denoted by $\sigma_{\eta;
Lm}$) participating the ultraspherical one, have either natural
$(\eta=+)$, or unnatural $(\eta=-)$ parities. In other words,
they transform with respect to the space inversion operation $P$
as

\begin{equation}
P\sigma_{\eta;Lm}=\eta e^{i\pi L}\sigma_{\eta;L-m},\quad
L=0^\eta,
1^{-\eta},...(\sigma-1)^{-\eta}, \quad m=-L,...,L
\label{5}
\end{equation}
In coupling a Dirac spinor, $\{1/2,0\}\otimes \{0,1/2\}$, to the
Coulomb multiplets $\{k/2,k/2\}$ from above, the spin $(J)$ and
parity $(P)$ quantum numbers of the BR's are created as

\begin{equation}
J^P=1/2^\eta, 1/2^{-\eta}, 3/2^{-\eta},...,(k+1/2)^{-\eta},\quad k=\sigma-1
\label{6}
\end{equation}
The RS fields are finite-dimensional nonunitary representations of
the Lorentz group which, in being described by totally symmetric
traceless rank-k Lorentz tensors with Dirac spinor components,
$\psi_{\mu 1\mu 2...\mu k}$, have the appealing property that
spinorial and four-vector indices are separated. They satisfy the
Dirac equation (DE) according to:

\begin{equation}
(i\partial\gamma -M)\psi_{\mu 1\mu 2...\mu k}=0
\label{7}
\end{equation}
The author claimed that, in terms of the notations introduced
above, {\it all} reported baryons with masses below 2500 MeV, are
completely accommodated by the RS fields $\psi_\mu$, $\psi_{\mu
1\mu 2\mu 3}$, and $\psi_{\mu 1\mu 2...\mu5}$, having states of
the highest spin $- 3/2^-$, $7/2^+$, and $11/2^+$, respectively
(see Figs.10-12 for 3D representation). In each one of the $N$,
$ \Delta$, $\Lambda$ and spectra, the RS cluster of lowest mass is
always $\psi_\mu$. For the nonstrange baryons, the $\psi_\mu$
cluster is followed by $\psi_{\mu 1\mu 2\mu 3}$, and $\psi_{\mu
1\mu 2...\mu 5}$, while for the $\Lambda$ hyperons a parity
doubling of the resonances starts above 1800 MeV. In the
following we will extend the notation of the RS clusters to
include isospin (I) according to $\sigma_{2I,\eta}$. For example,
the first $N$ cluster is denoted by $2_{1,+}$ , while $2_{3,+}$
and $2_{0,+}$ stand in turn for the corresponding $\Delta-$ and
$\Lambda-$hyperon ones. From Eqs.\ref{5} and \ref{6} follows that
the $2_{2I,+}$ clusters, where $I=1/2, 3/2$, and $0$, always unite
the first spin$-1/2^+$, $1/2^-$, and $3/2^-$ resonances.

Indeed, the relative $\pi N$ momentum $L$ takes for $l=0^+$ the
value $L=1^+$ and corresponds to the $P_{2I,1}$ state, while for
$l=1^-$ it takes the two values $L=0^-$ and $L=2^-$ describing in
turn the $S_{2I,1}$ and $D_{2I,3}$ resonances. The natural parity
of the first $O(4)$ harmonics reflects the arbitrary selection of
a {\it scalar vacuum} \cite{14} through the spontaneous breaking
of chiral symmetry. Therefore, up to three lowest $N$, $\Delta$,
and $\Lambda$ excitations, chiral symmetry is still in the
Nambu-Goldstone mode. The Fock space of the $2_{2I,+}$ clusters
will be denoted in the following by $F_+$. Note, that in this
context the first $P_{11}$ and $S_{11}$ states do not pair,
because their internal $L$ differ by one unit, instead of being
equal but of opposite parities. All the remaining $N$, $\Delta$
have been shown to belong to either $4_{2I,-}$, or $6_{2I,-}$.
They have been viewed to reside in a different Fock space, which
is built on top of a {\it pseudoscalar vacuum} \cite{14}.

For example, one finds all the seven $\Delta$'s \hskip 0.2 cm
$S_{31}, P_{31}, P_{33}, D_{33}, D_{35}, F_{35}$, and $F_{37}$
from the $4_{3,-}$ cluster to be squeezed within the narrow mass
region from 1900 MeV to 1950 MeV, while the $I=1/2$ resonances
paralleling them, of which only the $F_{17}$ state is still
"missing" from the data, are located around 1700 MeV (compare
Figs. 10,11).

In continuing by paralleling baryons from the third $N$ and
$\Delta$ clusters with $\sigma=6$, one finds in addition the four
states $H_{1,11}$, $P_{31}$, $P_{33}$, and $D_{33}$ with masses
above 2000 MeV to be "missing" for the completeness of the new
classification scheme. The $H_{1,11}$ state is needed to parallel
the well established $H_{3,11}$, while the $\Delta-$states
$P_{31}$, $P_{33}$, and $D_{33}$ are required as partners to the
(less established) $N_{1/2^+} (2100)$, $N_{3/2^+} (1900)$, and
$N_{3/2^-} (2080)$.

The degeneracy group of the $N$, $\Delta$ spectra found in
\cite{13} on the grounds of the successful RS classification is

\begin{equation}
SU(2)_I \times SU(3)_C \times O(1,3)_{LS}
\label{8}
\end{equation}

Traditionally, hadrons are classified in terms of
$SU(6)_{SF}\times O(3)_L$ multiplets. This classification is one
of the most important paradigms in hadron spectroscopy. According
to it, states like $N_{3/2^+}(1720)$, $N_{5/2^+}(1680)$,
$\Delta_{5/2^+}(1905)$, and $\Delta_{7/2^+}(1950)$, are viewed to
belong to a $56(2^+)$-plet, the $N_{1/2^+}(1710)$ excitation is
treated as a member of a $70(0^+)$-plet, while the negative parity
baryons $N_{1/2^-}(1535)$, $N_{3/2^-}(1520)$, $N_{1/2^-}(1650)$,
$N_{3/2^-}(1700)$, and $N_{5/2^-}(1675)$ are assigned to a
$70(1^-)$-plet. The above examples clearly illustrate that states
from Kirchbach's RS clusters separated by only few MeV, such as
$N_{5/2^-}(1675)$, $N_{5/2^+}(1680)$, and $N_{1/2^+}(1710)$ from
$4_{1^-}$, are distributed over three different $SU(6)_{SF}\times
O(3)_L$ representations, while on the other hand, resonances from
different RS "packages" separated by about 200 MeV, such as
$N_{3/2^-}(1520)$ from $2_{1^+}$, and $N_{3/2^-}(1700)$ from
$4_{1^-}$ are assigned to the same multiplet. This means that
$SU(6)_{SF}\times O(3)_L$ cannot be viewed as the degeneracy group
of baryon spectra. Further, the $SU(6)_{SF}\times O(3)_L$
multiplets appear approximately only "half-filled" by the reported
resonances. Several dozens states are "missing" for the
completeness of this classification scheme. As long as observed
and "missing" states are part of the same multiplets, they are
indistinguishable from the viewpoint of the underlying
$SU(6)_{SF}\times O(3)_L$ symmetry and there is no reason not to
believe to the observability of all states. On the contrary,
within the RS scheme, observed and "missing" states will fall
apart and be attributed to Lorentz multiplets of {\it different
space-time properties}.

The author also argued that the algebra of the degeneracy group
from Eq.\ref{8} is also partly the spectrum generating algebra.
Indeed, the reported mass averages of the resonances from the RS
multiplets with $L=1$, 3, and 5 are well described by means of the
following simple empirical recursive relation:
\begin{equation}
M_{\sigma^{\prime}}-M_\sigma=m_1\left({{1}\over{\sigma^2}}-{{1}\over{(\sigma^\prime)^2}}\right)+m_2{\left({{(\sigma^{\prime
2}-1)}\over{4}}-{{(\sigma^2-1)}\over{4}}\right)},
\label{9}
\end{equation}
where, again, $\sigma=k+1$.

The two mass parameters take the values $m_1=$600 MeV, and $m_2=$
70 MeV (so, $m_1\gg m_2$), respectively. The first term is the
typical difference between the energies of two single particle
states of principal quantum numbers $\sigma$, and
$\sigma^{\prime}$, respectively, occupied by a particle with mass
m, moving in a Coulomb-like potential of strength $\alpha_C$ with
$m_1=\alpha^2_Cm/2\hbar^2$. The term
\begin{equation}
{{\sigma^2-1}\over{2}}=k(k/2+1), with \quad k=\sigma-1,
\label{10}
\end{equation}
in Eq.\ref{9} is the generalization of the three-dimensional
$J(J+1)$ rule (with $J=k/2$) \hskip 0.2 cm to \hskip 0.2 cm four
\hskip 0.2 cm Euclidian \hskip 0.2 cm dimensions \hskip 0.2 cm and
describes \hskip 0.2 cm a generalized \hskip 0.2 cm O(4)
rotational \hskip 0.2 cm band. The parameter $1/m_2 = 2.82$ fm
corresponds to the moment of inertia $\tau= 2/5$ MR${^2}$ of some
"effective" rigid-body resonance with mass $M=1085$ MeV and a
radius $R=1.13$ fm. Therefore, the energy spectrum in Eq.\ref{9}
can be considered to emerge from a quark-C-hyperquark model with a
Coulomb-like potential $(H_{Coul})$ and a four-dimensional rigid
rotator $(T^{(4)}_{rot})$. The corresponding Hamiltonian $H^{QHM}$
that is diagonal in the basis of the O(4) harmonics is given by

\begin{equation}
H^{QHM} = H_{Coul} + T^{(4)}_{rot} =-\alpha_c/r+(1/2\tau) F^2
\label{11}
\end{equation}
Here, $(1/2\tau)F^2$ denotes a rigid rotator in four
Euclidian dimensions as associated with a {\it collective effect}
there. The author claimed: while the splitting between the
Coulomb-like states decreases with increasing $\sigma$, the
difference between the energies of the rotational states increases
linearly with $\sigma$ so that the net effect is an approximate
equidistancy of the baryon cluster levels.

Let's check this out with our modified CGI multiplets (Table 4).
As one can see from modified CGI mass differences, there is no
equidistancy in the baryon cluster levels. Instead the separation
is basically increasing with $J$ and $M$, while strong
fluctuations occurred. An average mass separation is 220 MeV.

So the author showed that $N^*$ and $\Delta^*$, instead of being
uniformly distributed in mass, as naively expected on the basis of
a 3q-Hilbert space without degeneracy, form well-pronounced {\it
spin-} and {\it parity-}clusters. The masses of the RS-clusters
and their spacing were shown to follow O(4) rotational bands
slightly modified by a Balmer-like term (see also \cite{15}).

We still have to point to some inconsistencies in the Kirchbach's
model. Let's put $N, \Delta, \Lambda, \Sigma$ spectra on one plot
(see Fig.13). One can definitely witness quite strange features of
RS clusters: {\it first} $N^*$-cluster has 95 MeV width, {\it
second} $N^*$-cluster has 70 MeV width, and {\it third}
$N^*$-cluster has 350 MeV width. So the last $N^*$-cluster has a
huge width, which is incompatible with reasonable definition of
the cluster.

{\it First} $\Delta^*$-cluster has 130 MeV width, {\it second}
$\Delta^*$-cluster has 50 MeV width, and {\it third}
$\Delta^*$-cluster has 420 MeV width. Again, the last
$\Delta^*$-cluster has a huge width, which is incompatible with
reasonable cluster definition. $\Lambda$-hyperons have one cluster
in this scheme at low masses, but have big 195 MeV width, which
hardly fits cluster definition.

{\it First} $\Sigma^*$-cluster has a reasonable 80 MeV width.
{\it Second} $\Sigma^*$-cluster and {\it third} $\Sigma^*$-cluster
have 360 and 370 MeV widths correspondingly, which make them bad
candidates for the realistic clusters. But that's not all: second
and third $\Sigma^*$-cluster have a {\it strong overlap} for a 110
MeV, making a $\Sigma$-sector a real failure of the Kirchbach's
model.

If we will try to merge $N-\Delta-\Lambda-\Sigma$ spectra onto
one plot of the concentric rings, like in our model, nothing
similar will happened. RS-rings will mostly fuse into a couple of
very broad rings, with practically no spacings between them (see
Fig.14). It is interesting to find some correspondence between
RS-clusters and our clusters (see Fig.13). We see that our
$1N(1650-1720)$ completely coincides with $2N^*(1650-1720)$ RS
from Kirchbach.

Third $N^*$ RS cluster $\psi_{\mu \eta \rho \tau \nu}(1900
-2250)$ covers completely our $2N$, $3N$, and $4N$ clusters. In
other words our $2,3,4$-$N$-clusters are inserted exactly into
$\psi_{\mu \eta \rho \tau \nu}$.

First $\Delta^*$ RS cluster $\psi_\mu(1620-1750)$ covers totally
second $N^*$ RS cluster $\psi_{\mu \eta \rho}(1650 -1720)$, and
our $1N-$, and $1\Sigma-$ clusters.

Second $\Delta^*$ RS cluster $\psi_{\mu \eta \rho}(1900 -1950)$ is
identical to our $1\Delta$.

Third $\Delta^*$ RS, $\psi_{\mu \eta \rho \tau \nu}(2000 -2420)$
cover totally our $3N$, $4N$, $3\Sigma$, and $2\Delta$.

Third $N^*$ RS and third $\Sigma^*$ RS are totally corresponding
to each other: $3N^*(1900-2250)$ is exactly inserted into
$3\Sigma^*(1880-2250)$.

\vskip 0.6 cm

{\bf \ \ \ \ \ \ \ \ \ \ \ \ \ \ \ \ \ \ \ \ \ \ \ \ \ \ \ \ \ \ \
\ \ \ \ {III. CONCLUSIONS}}

\vskip 0.1 cm

In this paper we analyze four state-of-the-art models of hadronic
structure. Chiral symmetry plays an eminent role in the formation
of the spectra. This symmetry is broken only by the very small
$u,d$ quark masses. However, this symmetry is not directly
observed in the world - it is {\it hidden}, i.e. spontaneously
broken.

The Cohen-Glozman model with their chiral quartets, is based on
the idea of chiral symmetry breaking on a vacuum level and then
full restoration of it high in the spectrum, precisely from 2 GeV.
The authors claim that upwards from 2 GeV parity doublets and
chiral quartets arise - and the precise chiral symmetry
restoration is responsible for the {\it genesis} of these doublets
and quartets.

Our results together with the Klempt model show that parity
doublets and chiral multiplets are indeed observed, but:

1) This start to happened far below 2 GeV.

2) The multiplet's width is growing as we are moving higher and
higher in mass and $J$.

3) Standard GC chiral multiplets are not properly filled from
$J=11/2$.

4) The experimental chiral multiplet structure could even better
be described in $LS$ multiplet sheme ( or, using $SU(6)_{SF}\times
O(3)_L$ group).

So, in fact one can see the sprouts of chiral symmetry
restoration above 2 GeV, but there are definitely a few
mechanisms, which are responsible for the parity doublets and
chiral multiplets formation in the current resonance energy
region.

Klempt's model has a few nice features - formation of the
$LS$ multiplets, reminiscent of the experimental clusterings (see
Figs.2-4), and introduction of the {\it solitaire states} and an
attraction between them. On the other hand, neglecting of the
{\it \textbf{LS}} coupling and mixing of the different orbitals in
baryon WF sometimes lead to wrong results (see the discussion
above on $N_{{3/2}^+}(1720)$ and other similar states).

Kirchbach aspires to have discovered a {\it degeneracy group} of
the baryonic spectra: $SU(2)_I\times SU(3)_C\times O(1,3)_{LS}$.
This group was inspired by H\"{o}ehler discovery of the $N$,
$\Delta$ spectra clustering phenomenon, i.e. - H\"{o}ehler poles,
or H\"{o}ehler clusters. All resonances are classified as
Rarita-Schwinger multispinors, and spin-parity multiplets arise.
To the advantage of this model we have to attribute the prediction
of both existing and missing BR's, belonging to Lorentz multiplets
of {\it different} space-time properties. Nevertheless we have to
admit that defined this way, RS-clusters are getting very broad at
already low masses. What's even more - second and third
$\Sigma$-clusters have a strong overlap, making RS-scheme
non-applicable in $\Sigma$-sector.

We have considered baryonic
structure, based on three different groups:

$$\cases{SU(2)_R\times SU(2)_L \quad (and, SU(3)_R\times SU(3)_L )\cr
O(3)_L\times SU(6)_{SF}\cr
SU(2)_I\times O(1, 3)_{LS}\times SU(3)_C}$$
All of these groups predict clustering patterns in $N, \Delta,
\Lambda, \Sigma$ spectra, but these clusters are {\it different}.
Even the topology of resulting towers is different and reflected
in regular and flipped pyramidal structures (see Fig.15).

Klempt's and Inopin multiplets exhibit "rings structure", or
"solar-type structure" with exactly 10 rings - copies of the Solar
planets (see Fig.16). The distances between the rings vary - so
does this happen in the Solar system as well.

We certainly hardly could give preference to any of the above
models - each has its own bold features, which are absent in the
others. We have also uncovered a number of interesting symmetries
and multiplet structures in mesonic sector, but this story will be
unfolded in a separate paper.

Finally we have to mention the model of Hatsuda et al \cite{16}.
This model has the spirit of CG model, with interesting "mirror
ansatz". Unfortunately, the authors consider only low-lying
$N-\Delta$ quartets, which makes their model of a limited value.
\vskip0.3 cm \textbf{Acknowledgement}: Author is very grateful
to C.F. Diether III for proofreading and making a website for my
color plots.


\begin{thebibliography}{99}
\bibitem{1}T.P. Cheng, L.F. Li, {\it Gauge Theory of Elementary Particle Physics},
Clarendon Press, Oxford, 1984.

\bibitem{2}G. H\"{o}ehler, In: {\it Physics with GeV -particle beams}, eds. H. Machner
and K. Sistemich (World Scientific, Singapore) 1995, p.198.

\bibitem{3}F. Iachello, Phys.Rev.Lett., v.62 (1989), 2440.
\bibitem{4}T.D. Cohen, L.Ya. Glozman, Int.J.Mod.Phys.A, vol.17, (2002),
1327.

\bibitem{5} Particle Data Group, K. Hagiwara et al., Phys.Rev.D 66, 010001
(2002).

\bibitem{6} A.E. Inopin, E.V. Inopin, Sov.J.Nucl.Phys. 53(2), (1991), 351.

\bibitem{7}A.E. Inopin, G.S. Sharov, Phys.Rev.D 63, 054023 (2001).

\bibitem{8}E. Klempt, nucl-ex/0203002; Phys.Rev.C 66, 058201 (2002);
hep-ph/0212241.

\bibitem{9}A.E. Inopin, hep-ph/0012248.

\bibitem{10}A.E. Inopin, hep-ph/0110160 (review).

\bibitem{11}F. Iachello et al., Ann.Phys. v.284 (2000), 89.

\bibitem{12}N. Isgur, S.Capstick, Phys.Rev.D 34 (1986), 2809.

\bibitem{13}M. Kirchbach, Int.J.Mod.Phys. A15 (2000), 1435.

\bibitem{14}P.W. Milonni, {\it The Quantum Vacuum. An Introduction to Quantum
Electrodynamics}. Academic Press, 1994, 522p.

\bibitem{15}M.Kirchbach, M. Moshinsky, Yu.F. Smirnov, Phys. Rev.
D64, 114005 (2001).
\bibitem{16}T. Hatsuda, D. Jido, T. Kunihiro, Phys.Rev.Lett. v.84 (2000), 3252.

\end{thebibliography}
\end{document}